\documentclass[a4paper]{article}
\usepackage{graphicx}
\usepackage{algorithm}

\makeatletter
\renewcommand{\ALG@name}{Procedure} 
\makeatother

\usepackage{multirow}
\usepackage{multicol}
\usepackage{xcolor}
\usepackage{amsthm,amsmath}
\usepackage{algorithm,algorithmic}
\usepackage[utf8]{inputenc} 

\def\includegraphics{}

\title{Burrows Wheeler Transform on a Large Scale: Algorithms Implemented in Apache Spark}

\author{Ylenia Galluzzo \and Raffaele Giancarlo \and Mario Randazzo \and Simona E. Rombo}

\date{

	Department of Mathematics and Computer Science, University of Palermo, Palermo, Italy \\ 
	\texttt{\{ylenia.galluzzo, mario.randazzo\}@community.unipa.it}\\[2ex]%
	\texttt{\{raffaele.giancarlo, simona.rombo\}@unipa.it}\\[2ex]
	
}

\begin{document}
	\maketitle
\begin{abstract} 
  With the rapid growth of Next Generation Sequencing (NGS) technologies, large amounts of ``omics'' data are daily collected and need to be processed. Indexing and compressing large sequences datasets are some of the most important tasks in this context. Here we propose algorithms for the computation of Burrows Wheeler transform relying on Big Data technologies, i.e., Apache Spark and Hadoop. Our algorithms are the first ones that distribute the index computation and not only the input dataset, allowing to fully benefit of the available cloud resources.  

\end{abstract}




\section*{Introduction}
The enormous amount of data produced by the Next-Generation Sequencing (NGS) technologies opens the way for a more comprehensive characterization of mechanisms at the molecular level, which are at the basis of cellular life and may have a role in the occurrence and progress of disorders and diseases. This may help to answer to fundamental questions for biological and clinical research, such as how the interactions between cellular components and the chromatin structure may affect gene activity, or to what extent complex diseases such as diabetes or cancer may involve specific (epi)genomic traits. Indexing NGS data is an important problem in this context \cite{ceri}. 

In particular, an index is a data structure that enables efficient retrieval of stored objects. Indexing strategies used in NGS allow  space-efficient storage of biological sequences in a  {\it full-text index} that enables fast querying, in order to return exact or approximate string matches. Popular full-text index data structures include variants of suffix arrays \cite{AbouelhodaKO04}, FM-index based on the Burrows–Wheeler transform (BWT) and some auxiliary tables \cite{FerraginaM05}, and hash tables \cite{pone}. The choice of a specific index structure is often a trade-off between query speed and memory consumption. For example, hash tables can be very fast but their memory footprint is sometimes prohibitive for large string collections \cite{schmit17}. 

Here we address the problem of computing BWT in the distributed, exploiting Big Data technologies such as Apache Spark \cite{spark}. In particular, previous research has been proposed on the BWT computation in a MapReduce \cite{DeanG10} fashion based on Apache Hadoop \cite{menon11}. The use of Spark and Hadoop together, as proposed here, have shown to notably improve the performance in several application contexts, due to the optimal exploitation of both memory and cloud. Another available tool relying on Hadoop and BWT computation is BigBWA \cite{DAbuinPPA15}. However, the BigBWA parallelism is intended only to split the input sequences and then apply another existing framework, i.e., BWA \cite{LiD09}, in order to align them via BWT. Therefore, the BWT computation is not based itself on Big Data technologies in BigBWA.

We propose two algorithms for the BWT computation that fully exploit parallelism afforded by a cloud computing environment, combining advantages of MapReduce paradigm and Spark Resilient Distributed Datasets (RDD). 
Validation results obtained on real biological datasets, including genomic and proteomic data, are provided, showing that our approach improves the performance for BWT computation with respect to its competitors.

\subsection*{Preliminaries}

Let $S$ be a string of $n$ characters defined on the alphabet $\Sigma$. We denote by $S(i)$ the $i$-th character in $S$ and by $S_i$ its $i$-th suffix. We recall the following basic notions.

\paragraph{BWT} The Burrows-Wheeler transform of $S$ is useful in order to rearrange it into runs of similar characters. This may have advantages both for indexing and for compressing more efficiently $S$. The BWT applied to $S$ returns: 
\begin{itemize}
    \item a permutation $bwt(S)$ of $S$, obtained by sorting all its circular shifts in lexicographic order, and then extracting the last column;
    \item the index (0-based) $I$ of the row containing the original string $S$.
\end{itemize} 

\noindent Among the most important properties of BWT, it is reversible. Figure \ref{fig::bwt} shows an example of BWT for the string $S$=$BANANA\$$. In particular, $bwt(S)=BNN\$AAA$, and $I=3$.

\paragraph{Suffix Array} 
The suffix array $SA$ of $S$ is defined as an array of integers providing the starting positions of suffixes of $S$ in lexicographical order. Therefore, an entry $SA[i]$ contains the starting position of the $i$-th suffix in $S$ among those in lexicographic order. Figure \ref{fig::sa} shows the Suffix Array for the same example of BWT.

\paragraph{Inverse Suffix Array} 
Inverse Suffix Array of $S$, $ISA[i] = j$ means that the rank of the suffix $i$ is $j$, i.e., $SA[j] = i$.

\section*{Implementation}

We use the notation $[i,j]$ for denoting the set $\{i,\dots,j\}$.
Let $S \in \Sigma^*$ a string of length $n$. For $i \in [0, n-1]$, let $S_i$ denote the suffix of $S$ starting  in position $i$ and let $S_{i,j}$ denote the sub-string $S[i]S[i+1]\dots S[j]$. In order for the bwt calculation to be easily reversed we assume that the string S always ends with a \$ sentinel character, i.e., the smallest character in $\Sigma$. In addition, let's assume that for $S[i]= \$$ for $i > n$.

\noindent In the following, the two algorithms proposed here are described in detail.

\subsection*{Sorting based MapReduce (SMR)}

The first phase of the algorithm SMR is suffix partitioning. 
The goal is to partition the set of possible suffixes into sub-sets $K_{1},..., K_{r},$ where $r$ is a positive integer representing the desired number of partitions (i.e. the number of nodes within the cluster). The partitioning has to  comply with the following
property.

\noindent \textit{{\bf Property 1} For each pair $K_{i},K_{j}$, all suffixes in $K_{i}$ are lower or greater  in lexicographical order than all those ones in $K_{j}$.}

To this aim, suffixes are discriminated by their first $k$ characters (i.e., $k$-mers in the following). Here, $k$ is a positive integer chosen in advance. The algorithm SMR maps each suffix $S_{i}$ of $S$ in a key-value tuple {\it (k-mer, i)} (\textsc{Procedure 1}). Therefore, suffixes are partitioned according to the key, and then the tuples are sorted with respect to this key to maintain the order of the partitions.

\paragraph{Partitioning.} A fairly important issue is how to implement partitioning in practice. An efficient technique is to sample the set of keys and then determine ranges based on the desired number of partitions. Then the set of keys is partitioned according to the determined ranges, thus that the partitions are balanced. In this way, it is possible to partition and sort at the same time.
In \cite{menon11} this technique is used and optimized for the case of genomics sequences. It is also provided by the Spark framework \cite{zaharia2012resilient} through the RangePartitioner functionality \footnote{\texttt{https://spark.apache.org/docs/2.3.0/api/java/org/apache/spark/RangePartitioner.html}}.
\noindent \\
\textit{Example:}\\  

\-\hspace{1.5cm}$S = $\texttt{ C\quad A\quad T\quad T\quad A\quad T\quad T\quad A\quad G\quad G\quad A} \\ 
\-\hspace{2.1cm}\texttt{ 0\quad 1\quad 2\quad 3\quad 4\quad 5\quad 6\quad 7\quad 8\quad 9\quad 10}\\
\-\hspace{2.1cm}\texttt{ \quad\quad\quad *\quad * \quad\quad *\quad * }\\\\
\-\hspace{1.0cm} For $k = 3$ \\
\-\hspace{1.2cm}
\begin{tabular}{rr}
	\rule[-2ex]{0pt}{4ex} & Sorted partitions\\
	   (\texttt{CAT},0) &  \{(\texttt{\$\$\$},11),(\texttt{A\$\$},10) \} \\
	   (\texttt{ATT},1) &   \{(\texttt{ATT},1),(\texttt{ATT},4),(\texttt{AGG},7) \} \\
	   (\texttt{TTA},2) &   \{(\texttt{CAT},0),(\texttt{GA\$},9),(\texttt{GGA},8) \} \\
	   (\texttt{TAT},3) &    \{(\texttt{TTA},2),(\texttt{TAT},3),(\texttt{TTA},5),  \} \\ (\texttt{ATT},4)  \\
	   (\texttt{TTA},5)  \\
	   (\texttt{TAG},6)  \\
	   (\texttt{AGG},7)  \\
	   (\texttt{GGA},8)  \\
	   (\texttt{GA\$},9)  \\
	   (\texttt{A\$\$},10)  \\
	   (\texttt{\$\$\$},11)  \\
\end{tabular} 
\begin{algorithm}[H]
\caption{Preparation to Partitioning}
\begin{algorithmic}[1]
\STATE\textbf{procedure }{\textsc{MAP}}{(\textit{$S_{i}$})}
  \STATE \textbf{return} ($k$-mer of $S_{i}$, $i$)
\end{algorithmic}
\label{alg:alg_iterative_radix_sort_partitioning}
\end{algorithm}
The second stage consists of completing the work by ordering suffixes partition by partition (see \textsc{Procedure 2}). The task is therefore to collect the suffix indexes $p$ in the partition and to produce by a single procedure what is here called \textit{Partial SA}.
The idea is to reduce the problem of calculating the partial SA $SA_{p}$ to the calculation of another SA, $SA_{t}$, that refers to a new string $T$ built from $p$. The order of suffixes in $T$ implicitly defines the order of suffixes indexed from $p$ in $S$. This idea is behind the recursive step used in the DC3 \cite{karkkainen2006linear} algorithm.
\begin{algorithm}[H]
\caption{Partial Suffix Arrays Computation}
\begin{algorithmic}[1]
\STATE\textbf{procedure }{\textsc{CALCULATE PARTIAL SA}}{(\textit{p})}
  \STATE Calculate $l_{max}$ the maximum distance between two elements in $p$ \\
  \STATE Generate $\mathcal{L}$ the list of ${S[p[i], l_{max}]}$ for each $i \in p$  \\
  \STATE Sort $\mathcal{L}$ using \textsc{AlgorithmX} \\
  \STATE \textbf{return} \textit{$SA_{p}$}
\end{algorithmic}
\label{alg:alg_iterative_radix_sort}
\end{algorithm}

Two different variants of SMR are considered: SMR$_r$, such that \textsc{AlgorithmX} is the Radix Sort, and SMR$_t$, if it is the Timsort.

\subsection*{Prefix Doubling Algorithm (PDA)}
The more crucial aspect for the BWT computation considered here is the calculation of the Suffix Array of the input string. Indeed, BWT can be calculated from the Suffix Array in a MapReduce fashion via join operation. Therefore, the algorithm proposed for the computation of the Suffix Array, based on the idea of \textit{prefix doubling} inspired by \cite{FlickA15}, is described below (see \textsc{Procedures 3} and \textsc{4}).     

\paragraph{Input:} 
Let $S$ be a string of length $n$, the input tuples set is:
$$\text{Input} = \{ (\texttt{null}, S(i)) : i = 1,\dots,n)\}$$

\paragraph{Output:} A set of tuples of the form $(i, r)$, where $i$ is the index of a suffix in $S$ and $r$ is its rank (i.e., its position in the list of the sorted suffixes). In the literature this is referred to as the ISA. For our purpose, the resulting output is inverted in order to obtain the Suffix Array of $S$ and, then, its BWT.

\begin{algorithm}[H]
\caption{Sketch Algorithm Iterative with Prefix Doubling}
\begin{algorithmic}[1]

\STATE\textbf{procedure }{\textsc{CALCULATEISA}}{(\textit{S})}{
  \STATE $Input = \{(null, S(i)): i = 1,...,n\}$ \\
  \STATE Initialize set \textit{ISA} with \textit{Input} \\
  \FOR{($k \gets 0$\ to\ $\lceil log_{2}n \rceil$)}{
        \STATE Apply the operation of \textit{Shifing} to \textit{ISA} obtaining two sets\\
        \STATE Join the two sets obtained with the operation of \textit{Pairing}
        \STATE Update \textit{ISA} by calling \textsc{RE-RANKING}
        }
       \ENDFOR \\
     \STATE \textbf{return} \textit{ISA}
  }
\end{algorithmic}
\label{alg:alg_iterative_prefix_doubling}
\end{algorithm}

\begin{algorithm}[H]
\caption{ Re-ranking}
\begin{algorithmic}[1]

\STATE\textbf{procedure }{\textsc{RE-RANKING}}{(\textit{Pairs})}{
  \STATE Sort tuple of pairs in Pairs by value  \\
  \FORALL{($(i,(r_{a1}, r_{a2})) \in Pairs$)}{
        \STATE $Let (j,(r_{b1}, r_{b2}))$ the previous pair to $(i,(r_{a1}, r_{a2}))$\\
        \IF{$(r_{a1}, r_{a2}) = (r_{b1}, r_{b2})$}
          \STATE $r_{new} = j$
        \ELSE
          \STATE Assign to $r_{new}$ the position in the sorted set \textit{Pairs}  of $(i,(r_{a1}, r_{a2}))$
        \ENDIF
        \STATE Update tuple $(i,r)$ in \textit{ISA} with tuple $(i,r_{new})$ 
        }
       \ENDFOR \\
  }
\end{algorithmic}
\label{alg:re_ranking}
\end{algorithm}

\paragraph{Initialization.} The first step is starting from the Input set and initialize the set of tuples $(i, r)$, as described in the previous paragraph. In this phase, the rank is calculated by the first character of the suffix. In particular, let $Occ(c)$ be the number of occurrences of the character lexicographically smaller of $c$ in the string $S$, then the rank of the suffix $i$ can be determined as $Occ(S(i))$.\\
In a MapReduce fashion, this can be accomplished by first counting the occurrences of each character in $S$, and then computing the cumulative sum $Occ$ on the sorted counts. The map and reduce steps are:
$$\texttt{map: } (\texttt{null}, S(i)) \rightarrow (S(i), 1)$$
$$\texttt{reduce: } (c, \texttt{list}[1,1,\dots,1]) \rightarrow (c, \text{sum of ones}) $$
From this $Occ$ is calculated locally by collecting the result.\\
The $ISA$ set can be then initialized with the following map step:
$$\texttt{map: } (\texttt{null}, S(i)) \rightarrow (i, occ(S(i)))$$

\paragraph{ISA Extending.} The next step is to extend each rank contained in the initialized $ISA$ by the whole suffix. Here we use a technique called \textit{Prefix Doubling} which is based on the following statement:
\begin{center}
	\textit{Given that the suffixes of a string are already sorted by their prefix of length $h$, we can deduce their ordering by their prefix of length $2h$.}
\end{center}
Given two suffixes $S_i$ and $S_j$ with an identical prefix of length $h$, we can deduce their sorting by comparing the order of the suffixes $S_{i+h}$ and $S_{j+h}$. Thus the idea is to pair, for each suffix $S_i$, its rank with the rank of the suffix $S_{i+h}$ (i.e., $(ISA[i]), ISA[i+h])$) and sort all these pairs in order to obtain the sorting by the prefix of length $2h$. Indeed, an iteration double the prefix, since the longest suffix has size $n$, all suffixes will be sorted after at most $\log_2(n)$ iterations.

\paragraph{Shifting and Paring.} To implement the above idea in a MapReduce fashion, we apply the two considered map steps to the latest $ISA$ calculated to obtain two different sets:
$$\texttt{map: } (i, r) \rightarrow (i, (r, 0))$$
$$\texttt{map: } (i, r) \rightarrow (i - 2^k, (-r, 0))$$
where $k$ is the number of the iterations minus one. 
The indices of rank are shifted this way, then the rank is paired by a reduce step. It is worth noticing that a negative number is used to denote a shifted rank, and the value is mapped as a tuple with a zero term in order to consider the ranks shifted that overflow the string length.\\
The union of the two obtained sets is considered and all tuples with a negative key are discarded (the corresponding ranks do not pair with any other rank in the set). The following reduce step is applied to the union:
$$\texttt{reduce: } (i, \texttt{list}[(r1,0), (r2,0)]) \rightarrow (i, (r1, -r2))$$
where $r2$ is the rank shifted. Some ranks may occur that are not reduced due to the unique key. These ranks overflow the length of $S$ and remain paired with zero. We denote the final set derived from this phase by $Pairs$.

\paragraph{Re-Ranking.} Our purpose is to extend the previous rank with a new rank, obtained by considering the prefix doubled. Therefore, we compute the new rank according to the tuple in $Pairs$ as follows: firs we sort all tuples by value, then we compare each tuple at position $i$ (after sorting) with the one in position $i-1$. If they are equal, the new rank is equal to the rank of the previous tuple, otherwise the new rank is $i$. Finally, a new ISA set with rank extended is obtained, and the procedure is iterated on it again.
All operations described above can be achieved also in a distributed manner:
\begin{itemize}
    \item For the sorting operation, a certain number of partitions can be identified by range into roughly equal ranges the elements in the set (the ranges can be determined by sampling the data). Then for each partition a sorting algorithm is applied that sort each partition locally. This is easily provided by the framework Apache Spark. 
    \item In order to compute the new rank, the partition identified previously is considered and the procedure above is applied locally, as described before, using the length of the partition and the offset (i.e., the number of elements in the previous partition) for computing the position of the tuples.
\end{itemize}

\subsubsection*{Example}
Let S = \textit{BANANA\$} be the input string of length $n = 7$. The input pairs are:  
\begin{equation*}
    \begin{gathered}
        \text{Input} = \{ (\texttt{null}, B), (\texttt{null}, A), (\texttt{null}, N),\\ (\texttt{null}, A), (\texttt{null}, N), (\texttt{null}, A), (\texttt{null}, \$) \}
    \end{gathered}
\end{equation*}
As for $Occ(c)$, it is shown in Table \ref{tab::occ}.

\noindent After the initialization, the initial ISA set is:
\begin{equation*}
    \begin{gathered}
        \text{ISA} = \{ (0, 3), (1, 0), (2, 4), (3, 0),\\ (4, 4), (5, 0),(6, 6)\}
    \end{gathered}
\end{equation*}
After the first iteration, the shifted tuples are:
\begin{equation*}
    \begin{gathered}
        \text{Shifted} = \{(-1, (-3, 0)), (0, (0, 0)), (1, (-4, 0)),\\ (2, (0, 0)), (3, (-4, 0)), (4, (0, 0)),(5, (-6, 0))\}
    \end{gathered}
\end{equation*}
After the the pairing we obtain the set:
\begin{equation*}
    \begin{gathered}
        \text{Pairs=}\{(0, (3, 0)), (1, (0, 4)), (2, (4, 0), (3, (0, 4),\\ (4, (4, 0), (5, (0, 6), (6, (6, 0)\} 
    \end{gathered}
\end{equation*}
Finally, we sort by value and we re-rank the indices. Then the new ISA is:
\begin{equation*}
    \begin{gathered}
        \text{ISA} = \{ (0, 3), (1, 1), (2, 4), (3, 1),\\ (4, 4), (5, 0),(6, 6)\}
    \end{gathered}
\end{equation*}
We observe that the only rank updated in this iteration is the one with index $5$, indeed shifting by $1$ it is possible to distinguish among the prefixes $AN$, $AN$ and $A\$$ corresponding to the suffixes $S_1$, $S_3$ and $S_5$.

\section*{Results}

The presented algorithms have been evaluated on real datasets taken from the Pizza$\&$Chili website \cite{pizzachili}, where a set of text collections of various types and sizes are available to test experimentally compressed indexes. In particular, the text collections stored on this website have been selected to form a representative sample of different applications where indexed text searching might be useful. From this collection, we have chosen the following three datasets:

\begin{itemize}
    \item PROTEINS, containing a sequence of newline-separated protein sequences obtained from the Swissprot database. 
    \item DNA, a sequence of newline-separated gene DNA sequences obtained from files of the Gutenberg Project. 
    \item ENGLISH, the concatenation of English text files selected from collections of the Gutenberg Project. 
\end{itemize}



We have implemented in Apache Spark the algorithms described here and the basic approach proposed in \cite{menon11}, and we have run them on the GARR Cloud Platform. In particular, we have configured the cluster with $1$ master and $33$ slave nodes, each node with $6$ VCore, $32$ GB of RAM and $200$ GB for disk. We have used Apache Hadoop $3.1.3$ and Spark $2.3.4$.

Results are shown in Table \ref{tab:time-result} (when the running time was larger than $10$ hours it has not been reported). For the PROTEINS dataset, it has been considered the only first $25$ MB, the only first $100$ MB and the full dataset.

\subsection*{Discussion}
From the results of the experiments it is evident that the SMR version with Radix-Sort presents very long and impractical elaboration times. Although the implementation is realized in C language, in order to optimize as much as possible the calculation, this is still very expensive with input files of modest size. From a theoretical point of view this is easily explained by the fact that the algorithm has a computational complexity equal to $O(|p| - l_{max})$, where we recall that $p$ is the partition identified and $l_{max}$ is the maximum distance between two indices in $p$ considering also the final index. For partitions where the indices are not uniformly distributed, $l_{max}$ may become very large, causing very slow processing time. The SMR version with Timsort has better performance. However, it is not able to process very large files. In contrast with the two SMR variants, PDA is able to process all the datasets. This is what we expected due to the fact that it fully introduces parallelism in the computation of BWT, allowing to  benefit of cloud computing. 

\section*{Conclusion}
Two MapReduce algorithms for the implementation of a full-text index, that is, the Burrows Wheeler transform, have been proposed here. The algorithms have been implemented in Apache Spark and they have been validated on real datasets.

Among the various applications where an efficient and distributed implementation of BWT may be useful (e.g., data compression, pattern matching, etc.), we mention that searching for a suitable combination of Indexing and Machine Learning techniques has recently proved to be a promising issue \cite{GRAHAM,raff2019new,FerraginaV20}. Therefore, we plan to focus our future studies in this direction.




\section*{Funding}
PRIN research project ``Multicriteria Data Structures and Algorithms: from compressed  to learned indexes, and beyond'', grant n. 2017WR7SHH, funded by MIUR.

\noindent GNCS 2020 research project ``Algorithms, methods and software tools for knowledge discovery in the context of Precision Medicine'', funded by INDAM.


\section*{Availability of data and materials}
\texttt{https://github.com/MR6996/spark-bwt}

\section*{Competing interests}
The authors declare that they have no competing interests.





\bibliographystyle{bmc-mathphys} 

\newcommand{\BMCxmlcomment}[1]{}

\BMCxmlcomment{

<refgrp>

<bibl id="B1">
  <title><p>Indexing Next-Generation Sequencing data</p></title>
  <aug>
    <au><snm>Jalili</snm><fnm>V</fnm></au>
    <au><snm>Matteucci</snm><fnm>M</fnm></au>
    <au><snm>Masseroli</snm><fnm>M</fnm></au>
    <au><snm>Ceri</snm><fnm>S</fnm></au>
  </aug>
  <source>Inf. Sci.</source>
  <pubdate>2017</pubdate>
  <volume>384</volume>
  <fpage>90</fpage>
  <lpage>-109</lpage>
</bibl>

<bibl id="B2">
  <title><p>Replacing suffix trees with enhanced suffix arrays</p></title>
  <aug>
    <au><snm>Abouelhoda</snm><fnm>MI</fnm></au>
    <au><snm>Kurtz</snm><fnm>S</fnm></au>
    <au><snm>Ohlebusch</snm><fnm>E</fnm></au>
  </aug>
  <source>J. Discrete Algorithms</source>
  <pubdate>2004</pubdate>
  <volume>2</volume>
  <issue>1</issue>
  <fpage>53</fpage>
  <lpage>-86</lpage>
</bibl>

<bibl id="B3">
  <title><p>Indexing compressed text</p></title>
  <aug>
    <au><snm>Ferragina</snm><fnm>P</fnm></au>
    <au><snm>Manzini</snm><fnm>G</fnm></au>
  </aug>
  <source>J. {ACM}</source>
  <pubdate>2005</pubdate>
  <volume>52</volume>
  <issue>4</issue>
  <fpage>552</fpage>
  <lpage>-581</lpage>
</bibl>

<bibl id="B4">
  <title><p>MOSAIK: A Hash-Based Algorithm for Accurate Next-Generation
  Sequencing Short-Read Mapping</p></title>
  <aug>
    <au><snm>Lee</snm><fnm>WP</fnm></au>
    <au><snm>Stromberg</snm><fnm>MP</fnm></au>
    <au><snm>Ward</snm><fnm>A</fnm></au>
    <au><snm>Stewart</snm><fnm>C</fnm></au>
    <au><snm>Garrison</snm><fnm>EP</fnm></au>
    <au><snm>Marth</snm><fnm>GT</fnm></au>
  </aug>
  <source>PLoS One</source>
  <pubdate>2014</pubdate>
  <volume>9</volume>
  <issue>3</issue>
  <fpage>e90581</fpage>
</bibl>

<bibl id="B5">
  <title><p>Next-generation sequencing: big data meets high performance
  computing</p></title>
  <aug>
    <au><snm>Schmidt</snm><fnm>B</fnm></au>
    <au><snm>Hildebrandt</snm><fnm>A</fnm></au>
  </aug>
  <source>Drug Discovery Today</source>
  <pubdate>2017</pubdate>
  <volume>22</volume>
  <issue>4</issue>
  <fpage>712</fpage>
  <lpage>-717</lpage>
</bibl>

<bibl id="B6">
  <title><p>Resilient Distributed Datasets: A Fault-Tolerant Abstraction for
  In-Memory Cluster Computing</p></title>
  <aug>
    <au><snm>Zaharia</snm><fnm>M</fnm></au>
    <au><snm>Chowdhury</snm><fnm>M</fnm></au>
    <au><snm>Das</snm><fnm>T</fnm></au>
    <au><snm>Dave</snm><fnm>A</fnm></au>
    <au><snm>Ma</snm><fnm>J</fnm></au>
    <au><snm>McCauly</snm><fnm>M</fnm></au>
    <au><snm>Franklin</snm><fnm>MJ</fnm></au>
    <au><snm>Shenker</snm><fnm>S</fnm></au>
    <au><snm>Stoica</snm><fnm>I</fnm></au>
  </aug>
  <source>Presented as part of the 9th {USENIX} Symposium on Networked Systems
  Design and Implementation ({NSDI} 12)</source>
  <publisher>San Jose, CA</publisher>
  <pubdate>2012</pubdate>
  <fpage>15</fpage>
  <lpage>-28</lpage>
</bibl>

<bibl id="B7">
  <title><p>MapReduce: a flexible data processing tool</p></title>
  <aug>
    <au><snm>Dean</snm><fnm>J</fnm></au>
    <au><snm>Ghemawat</snm><fnm>S</fnm></au>
  </aug>
  <source>Commun. {ACM}</source>
  <pubdate>2010</pubdate>
  <volume>53</volume>
  <issue>1</issue>
  <fpage>72</fpage>
  <lpage>-77</lpage>
</bibl>

<bibl id="B8">
  <title><p>Rapid Parallel Genome Indexing with MapReduce</p></title>
  <aug>
    <au><snm>Menon</snm><fnm>RK</fnm></au>
    <au><snm>Bhat</snm><fnm>GP</fnm></au>
    <au><snm>Schatz</snm><fnm>MC</fnm></au>
  </aug>
  <source>Proceedings of the Second International Workshop on MapReduce and Its
  Applications</source>
  <publisher>New York, NY, USA: Association for Computing Machinery</publisher>
  <series><title><p>MapReduce ‚Äô11</p></title></series>
  <pubdate>2011</pubdate>
  <fpage>51</fpage>
  <lpage>-58</lpage>
</bibl>

<bibl id="B9">
  <title><p>BigBWA: approaching the Burrows-Wheeler aligner to Big Data
  technologies</p></title>
  <aug>
    <au><snm>Abu{\'{\i}}n</snm><fnm>JM</fnm></au>
    <au><snm>Pichel</snm><fnm>JC</fnm></au>
    <au><snm>Pena</snm><fnm>TF</fnm></au>
    <au><snm>Amigo</snm><fnm>J</fnm></au>
  </aug>
  <source>Bioinformatics</source>
  <pubdate>2015</pubdate>
  <volume>31</volume>
  <issue>24</issue>
  <fpage>4003</fpage>
  <lpage>-4005</lpage>
</bibl>

<bibl id="B10">
  <title><p>Fast and accurate short read alignment with Burrows-Wheeler
  transform</p></title>
  <aug>
    <au><snm>Li</snm><fnm>H</fnm></au>
    <au><snm>Durbin</snm><fnm>R</fnm></au>
  </aug>
  <source>Bioinformatics</source>
  <pubdate>2009</pubdate>
  <volume>25</volume>
  <issue>14</issue>
  <fpage>1754</fpage>
  <lpage>-1760</lpage>
</bibl>

<bibl id="B11">
  <title><p>Resilient distributed datasets: A fault-tolerant abstraction for
  in-memory cluster computing</p></title>
  <aug>
    <au><snm>Zaharia</snm><fnm>M</fnm></au>
    <au><snm>Chowdhury</snm><fnm>M</fnm></au>
    <au><snm>Das</snm><fnm>T</fnm></au>
    <au><snm>Dave</snm><fnm>A</fnm></au>
    <au><snm>Ma</snm><fnm>J</fnm></au>
    <au><snm>McCauly</snm><fnm>M</fnm></au>
    <au><snm>Franklin</snm><fnm>MJ</fnm></au>
    <au><snm>Shenker</snm><fnm>S</fnm></au>
    <au><snm>Stoica</snm><fnm>I</fnm></au>
  </aug>
  <source>9th $\{$USENIX$\}$ Symposium on Networked Systems Design and
  Implementation ($\{$NSDI$\}$ 12)</source>
  <pubdate>2012</pubdate>
  <fpage>15</fpage>
  <lpage>-28</lpage>
</bibl>

<bibl id="B12">
  <title><p>Linear work suffix array construction</p></title>
  <aug>
    <au><snm>K{\"a}rkk{\"a}inen</snm><fnm>J</fnm></au>
    <au><snm>Sanders</snm><fnm>P</fnm></au>
    <au><snm>Burkhardt</snm><fnm>S</fnm></au>
  </aug>
  <source>Journal of the ACM (JACM)</source>
  <publisher>ACM New York, NY, USA</publisher>
  <pubdate>2006</pubdate>
  <volume>53</volume>
  <issue>6</issue>
  <fpage>918</fpage>
  <lpage>-936</lpage>
</bibl>

<bibl id="B13">
  <title><p>Parallel distributed memory construction of suffix and longest
  common prefix arrays</p></title>
  <aug>
    <au><snm>Flick</snm><fnm>P</fnm></au>
    <au><snm>Aluru</snm><fnm>S</fnm></au>
  </aug>
  <source>Proceedings of the International Conference for High Performance
  Computing, Networking, Storage and Analysis, {SC} 2015, Austin, TX, USA,
  November 15-20, 2015</source>
  <pubdate>2015</pubdate>
  <fpage>16:1</fpage>
  <lpage>-16:10</lpage>
</bibl>

<bibl id="B14">
  <title><p>The Pizza and Chili corpus home page</p></title>
  <aug>
    <au><snm>Manzini</snm><fnm>G</fnm></au>
    <au><snm>Navarro</snm><fnm>G</fnm></au>
  </aug>
  <source>Web site: http://pizzachili. dcc. uchile</source>
  <pubdate>2007</pubdate>
</bibl>

<bibl id="B15">
  <title><p>On the internal correlations of protein sequences probed by
  non-alignment methods: Novel signatures for drug and antibody targets via the
  {B}urrows-{W}heeler Transform</p></title>
  <aug>
    <au><snm>Graham</snm><fnm>DJ</fnm></au>
    <au><snm>Robinson</snm><fnm>BP</fnm></au>
  </aug>
  <source>Chemometrics and Intelligent Laboratory Systems</source>
  <pubdate>2019</pubdate>
  <volume>193</volume>
  <fpage>103809</fpage>
</bibl>

<bibl id="B16">
  <title><p>A New Burrows Wheeler Transform Markov Distance</p></title>
  <aug>
    <au><snm>Raff</snm><fnm>E</fnm></au>
    <au><snm>Nicholas</snm><fnm>C</fnm></au>
    <au><snm>McLean</snm><fnm>M</fnm></au>
  </aug>
  <pubdate>2019</pubdate>
</bibl>

<bibl id="B17">
  <title><p>The PGM-index: a fully-dynamic compressed learned index with
  provable worst-case bounds</p></title>
  <aug>
    <au><snm>Ferragina</snm><fnm>P</fnm></au>
    <au><snm>Vinciguerra</snm><fnm>G</fnm></au>
  </aug>
  <source>Proc. {VLDB} Endow.</source>
  <pubdate>2020</pubdate>
  <volume>13</volume>
  <issue>8</issue>
  <fpage>1162</fpage>
  <lpage>-1175</lpage>
</bibl>

</refgrp>
} 





\section*{Figures}
\begin{figure}[h!]
    \centering
    \begin{tabular}{|c|c|}
    \hline
         {\bf All rotations of $S$} & {\bf Lexicographic sorting} \\
         \hline
         $BANANA\$$&  $ANANA\${\mathbf B}$\\
         $\$BANANA$& $ANA\$BA{\mathbf N}$\\
         $A\$BANAN$& $A\$BANA{\mathbf N}$\\
         $NA\$BANA$& $BANANA{\mathbf \$}$\\
         $ANA\$BAN$& $NANA\$B{\mathbf A}$\\
         $NANA\$BA$& $NA\$BAN{\mathbf A}$\\
         $ANANA\$B$& $\$BANAN{\mathbf A}$\\
         \hline
    \end{tabular}
    \caption{Example of BWT.}
    \label{fig::bwt}
\end{figure}

\begin{figure}[h!]
    \centering
    \begin{tabular}{|c|l|l|c|}
    \hline
         {\bf i} & {\bf Suffixes} & {\bf Sorted Suffixes}&  {\bf SA[i]}\\
         \hline
         $1$& $BANANA\$$ &$ANANA\$$&$2$\\
         $2$& $ANANA\$$&$ANA\$$&$4$\\
         $3$& $NANA\$$&$A\$$&$6$\\
         $4$& $ANA\$$&$BANANA\$$&$1$\\
         $5$& $NA\$$&$NANA\$$&$3$\\
         $6$& $A\$$&$NA\$$&$5$\\
         $7$& $\$$&$\$$&$7$\\
         \hline
    \end{tabular}
    \caption{Example of Suffix Array.}
    \label{fig::sa}
\end{figure}


\section*{Tables}
\begin{table}[h!]
\caption{Computation of $Occ(c)$.}
\begin{center}
	\begin{tabular}{|c|c|c|c|c|}
		\hline 
		$c$ & $A$ & $B$ & $N$ & $\$$ \\ 
		\hline 
		$Occ(c)$ & 0 & 3 & 4 & 6  \\ 
		\hline 
	\end{tabular} 
\end{center}
    \label{tab::occ}
\end{table}

\begin{table}[h!]
\caption{Performance comparison between the proposed algorithms.}
\begin{tabular}{|l|r|r|r|}
\hline
\multicolumn{1}{|c|}{\multirow{2}{*}{\textbf{Input}}} & \multicolumn{3}{c|}{\textbf{Time}}                                        \\ \cline{2-4} 
\multicolumn{1}{|c|}{}                                & \multicolumn{1}{l|}{SMR$_r$}  & \multicolumn{1}{l|}{SMR$_t$} & \multicolumn{1}{l|}{PDA}\\ \hline
Proteins.200MB (50 MB)                           & 26 minutes                       & 26 minutes           &    4.13 minutes        \\ \hline
Proteins.200MB (100 MB)                          & 1.20 hours                        & 56 minutes        &  7.09        minutes   \\ \hline
Proteins.200MB (Full)                            & $>$ 2.30 hours                   & 1.35 hours        &  18.55    minutes         \\ \hline
Dna.200MB (50 MB)                                & 9.80 minutes                       &  34 minutes             & 4.69   minutes \\ \hline
Dna.200MB (100 MB)                               &  1.20 hours                       &   59 minutes             & 7.94    minutes           \\ \hline
Dna.200MB (Full)                                 & $>$ 2.30 hours                         &  2.30 hours            & 21.25   minutes \\ \hline
English.1024MB                                   & -                                  & -                        & 2.22 hours \\ \hline
\end{tabular}
\label{tab:time-result}
\end{table}




\end{document}